\documentclass[twocolumn]{aastex631}

\usepackage{amsmath}
\usepackage{tensor}
\usepackage{cleveref}

\newcommand{\eps}[2]{\tensor[^{\epsilon}]{#1}{#2}}

\newcommand{\ii}{\mathrm{i}}
\newcommand{\ip}{e^{\ii P}}
\newcommand{\iq}{e^{\ii Q}}
\newcommand{\ir}{e^{\ii R}}
\newcommand{\id}{e^{\ii D}}

\newcommand{\fg}{\omega_{\mathrm{g}}}
\newcommand{\fe}{\omega_{\mathrm{e}}}

\newcommand{\intNk}{\int_{\mathcal{N}}d^{3}\mathcal{N}\left(k\right)\,}
\newcommand{\intNm}{\int_{\mathcal{N}}d^{3}\mathcal{N}\left(m\right)\,}
\newcommand{\intCl}{\int_{\mathcal{C}}d\mathcal{C}\left(l\right)\,}
\newcommand{\intV}{\int_{\mathcal{V}}d^{4}\mathcal{V}\left(k,l\right)\,}

\newcommand{\Xp}{X^{\mathrm{p}}}
\newcommand{\Xh}{X^{\mathrm{h}}}

\newcommand{\tXp}{\tilde{X}^{\mathrm{p}}}
\newcommand{\tXh}{\tilde{X}^{\mathrm{h}}}

\newcommand{\Jp}{J^{\mathrm{p}}}
\newcommand{\Jh}{J^{\mathrm{h}}}

\newcommand{\tJp}{\tilde{J}^{\mathrm{p}}}
\newcommand{\tJh}{\tilde{J}^{\mathrm{h}}}

\reportnum{CTPU-PTC-22-17}

\begin{document}

\title{Observation of Gravitational Waves by Invariants for Electromagnetic Waves}
\author[0000-0002-2692-7520]{Chan Park}
\email{iamparkchan@gmail.com}
\affiliation{Center for Theoretical Physics of the Universe, Institute for Basic Science (IBS), Daejeon, 34126, Republic of Korea}

\begin{abstract}
    We lay a theoretical foundation in the observation of gravitational waves (GWs) by electromagnetic waves (EMWs) performing a full electromagnetic analysis without any optical approximation. For that, the perturbation of plane EMWs is obtained by solving the perturbed Maxwell equation with GWs in the Minkowski background spacetime. In a GW detector using the EMWs, we propose to measure the electromagnetic invariants that are independent of the motion of the EMW receiver and whose perturbations are gauge-invariant. Imposing a physical boundary condition at EMW emitters, we have analytic results for the perturbations of invariants that can be measured in the EMW receiver. Finally, we show antenna patterns of the detector with monochromatic plane GWs and EMWs.
\end{abstract}

\section{Introduction}

Electromagnetic waves (EMWs) are widely applied in modern gravitational wave (GW) detections \citep{hellingsUpperLimitsIsotropic1983,bondInterferometerTechniquesGravitationalwave2016}, which have had great success
\citep{abbottGWTC1GravitationalWaveTransient2019,abbottGWTC2CompactBinary2021} and allow for promising new discoveries \citep{shannonGravitationalWavesBinary2015,babakEuropeanPulsarTiming2016,arzoumanianNANOGrav12Yr2020}. The detection of GWs by EMWs are based on the perturbational analysis of the EMWs in a curved spacetime with the GWs, which was initiated in \cite{estabrookResponseDopplerSpacecraft1975}, \cite{sazhinOpportunitiesDetectingUltralong1978}, and \cite{detweilerPulsarTimingMeasurements1979}. In the analysis, the governing equations for the EMWs are provided by geometrical optics that describes the EMWs as rays \citep[chap. 22.5]{misnerGravitation2017}. Then, the trajectory and propagation time of the ray are determined by the laws of geometrical optics with boundary conditions at both ends of the ray \citep{rakhmanovRoundtripTimePhoton2009}. From the analytic prediction of the ray's time delay induced by GWs and its measurements, we extract the information about the GWs.

Note that geometrical optics is applicable only when the wavelength of EMWs in an observer is much smaller than the radius of spacetime curvature. Although modern GW detectors are designed under this condition, we consider a detection of GWs by EMWs beyond geometrical optics. We expect that it can be utilized for upgrading current detectors or designing new detectors. To consider the perturbed EMWs in general, a full electromagnetic analysis is required. There have been approaches to directly solving the Maxwell equation for the perturbed EMWs \citep{caluraExactSolutionHomogeneous1999,mielingElectromagneticFieldGravitational2021,parkObservationGravitationalWaves2021,parkExtendingObservationalFrequency2022}. However, because their applications were limited to the domain of geometrical optics, they were only to verify the results already derived from geometrical optics or to re-derive in other ways.

In this paper, we will show unexplored new aspects of the perturbed EMWs beyond geometrical optics and establish a theoretical basis for their applications to GW observations. For that, we will obtain the perturbation of plane EMWs by solving the perturbed Maxwell equation with gauge conditions that simplify the equation drastically. The observation of GWs by the electromagnetic invariants and its advantages will be introduced. Finally, we will show the antenna patterns of the observation with monochromatic plane GWs and EMWs. The natural unit is used by setting $c=1$. The lowercase Latin letters starting with $a$ in tensor indices are abstract indices \citep[chap. 2.4]{wald1984general}.

\section{Perturbation}

Let $\mathcal{F}$ be a 5-dimensional manifold foliated by a one-parameter family of spacetimes $\mathcal{M}_{\epsilon}$ with a perturbation parameter $\epsilon$ such that $\mathcal{M}_{0}$ is the background spacetime. To define a perturbation in the background, we need a one-parameter group of diffeomorphisms $\phi_{\epsilon}:\mathcal{M}_{0}\rightarrow\mathcal{M}_{\epsilon}$. For a physical quantity $W$ in the perturbed spacetimes, its perturbed value $\eps{W}{}$ at the background is given by the pull-back from $\mathcal{M}_{\epsilon}$ to $\mathcal{M}_{0}$ as $\eps{W}{}=\phi^{*}_{-\epsilon}W$. When we consider a case of $W=O\left(\epsilon^{n}\right)$ where $n$ is a nonnegative integer, $\eps{W}{}$ is expanded at $\epsilon=0$ as
\begin{equation}
    \eps{W}{}=\epsilon^{n}\left\{Z+\epsilon\dot{Z}+O\left(\epsilon^{2}\right)\right\},
\end{equation}
where $Z=O\left(1\right)$ is the leading-order value, and $\dot{Z}=O\left(1\right)$ is its perturbation. The perturbation is provided by
\begin{equation}
    \dot{Z}=\mathcal{L}_{\upsilon}Z,   
\end{equation}
where $\mathcal{L}_{\upsilon}$ is the Lie derivative along $\upsilon$ that is the generator of $\phi_{\epsilon}$. This approach for perturbation, developed in \cite{stewart1974} and \cite{stewart1990}, has the advantage of being able to define the perturbation covariantly. 

\begin{figure}
    \centering
    \includegraphics[width=0.9\columnwidth]{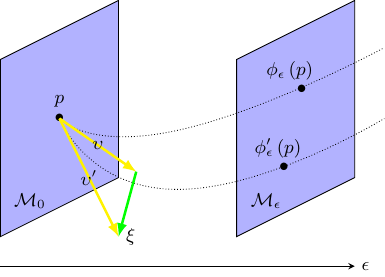}
    \caption{We show two different diffeomorphisms $\phi_{\epsilon}$ and $\phi'_{\epsilon}$ connecting $\mathcal{M}_{0}$ and $\mathcal{M}_{\epsilon}$ in the foliation of perturbed spacetimes. Orbits of each diffeomorphism are drawn with dotted line. Generators of diffeomorphisms $\upsilon$ and $\upsilon'$ (yellow vectors) are tangents for each orbit at $p\in\mathcal{M}_{0}$. The subtraction of generators $\xi=\upsilon'-\upsilon$ (green vector) is tangent to $\mathcal{M}_{0}$.}
    \label{fig:gauges_in_perturbation}
\end{figure}

Note that there are infinite ways to choose a map between $\mathcal{M}_{0}$ and $\mathcal{M}_{\epsilon}$. A change of map for perturbations from $\phi_{\epsilon}$ with $\upsilon$ to $\phi'_{\epsilon}$ with $\upsilon'$ is provided by
\begin{equation}
    \mathcal{L}_{\upsilon'}Z-\mathcal{L}_{\upsilon}Z=\mathcal{L}_{\xi}Z,\label{eq:gauge_transformation}
\end{equation}
where $\xi\equiv\upsilon'-\upsilon$ as depicted in \Cref{fig:gauges_in_perturbation}. Because $\xi\left(\epsilon\right)=\upsilon'\left(\epsilon\right)-\upsilon\left(\epsilon\right)=1-1=0$ from the definition, $\xi$ is tangent to $\mathcal{M}_{0}$ and $\mathcal{L}_{\xi}Z$ is evaluated in $\mathcal{M}_{0}$.  By formulating $\xi$ in \Cref{eq:gauge_transformation}, we can generate all possible perturbations. However, for a perturbation corresponding a physical reality, such ambiguities are not allowed. Therefore, for the physical perturbation, $\mathcal{L}_{\xi}Z$ should vanish for all $\xi$, which is possible only when $Z$ at the background is (i) zero, (ii) a constant scalar field, or (iii) a linear combination of products of identity tensor fields with constant coefficients as in \citet[chap 1.6]{stewart_1991}. In that sense, perturbations have gauges with the gauge transformation by \Cref{eq:gauge_transformation}, and the physical perturbation is gauge-invariant.

\section{Ansatz}

We set the background spacetime $\mathcal{M}_{0}$ to the Minkowski spacetime in which the Riemann curvature tensor vanishes. The perturbed metric is expanded into
\begin{equation}
    \eps{g}{_{ab}}=g_{ab}+\epsilon h_{ab}+O\left(\epsilon^{2}\right),
\end{equation}
where $g$ is the metric on the Minkowski spacetime, and $h$ is the metric perturbation. Assuming a perturbed stress-energy with second-order strength, $\eps{T}{_{ab}}=O\left(\epsilon^{2}\right)$, the perturbed Einstein equation becomes
\begin{equation}
    \nabla^{c}\nabla_{c}h_{ab}=0,\label{eq:wave_eq_of_h}
\end{equation}
with Lorenz and traceless gauge conditions given by
\begin{align}
    \nabla^{b}h_{ab}&=0,\label{eq:gc_Lorenz_of_h}
    \\\tensor{h}{^{a}_{a}}&=0,\label{eq:gc_traceless_of_h}
\end{align}
where $\nabla$ is the Levi-Civita connection associated with $g$ and raising or lowering indices are also obtained by $g$. We consider its wave solution as
\begin{equation}
    h_{ab}=\intNk\tilde{h}_{ab}\left(k\right)\ip,
\end{equation}
where $\mathcal{N}$ is the light cone given by $\mathcal{N}=\left\{k:k\cdot k=0\right\}-\left\{0\right\}$, dot($\cdot$) is the operator of a scalar product induced by $g$,
$\tilde{h}\,d^{3}\mathcal{N}$ is the infinitesimal amplitude, $\ii\equiv\sqrt{-1}$ is the imaginary unit, and $P$ is the phase such that $k^{a}=\nabla^{a}P$. Note that $\mathcal{N}=\mathcal{N}^{+}\cup\mathcal{N}^{-}$ where $\mathcal{N}^{+}$ and $\mathcal{N}^{-}$ are the future-directed and past-directed light cones, respectively. Because $h$ is real, the amplitudes of $\mathcal{N}^{+}$ and $\mathcal{N}^{-}$ have a correspondence given by $\tilde{h}\left(-k\right)=\tilde{h}^{*}\left(k\right)$. From the gauge conditions in \Cref{eq:gc_Lorenz_of_h,eq:gc_traceless_of_h}, we get
\begin{align}
    k^{b}\tilde{h}_{ab}\left(k\right)&=0,
    \\\tensor{\tilde{h}}{^{a}_{a}}\left(k\right)&=0,
\end{align}
for all $k\in\mathcal{N}$.

Let us consider a perturbed electromagnetic four-potential having first-order strength given by
\begin{equation}
    \eps{A}{_{a}}=\epsilon\left\{A_{a}+\epsilon X_{a}+O\left(\epsilon^{2}\right)\right\},
\end{equation}
where $A$ is the leading-order quantity, and $X$ is the perturbation. Because its stress-energy becomes $O\left(\epsilon^{2}\right)$, our choice is compatible to \Cref{eq:wave_eq_of_h}. Assuming an electromagnetic charge with $O\left(\epsilon^{3}\right)$ and choosing the Lorenz gauge for $\eps{A}{}$, the perturbed Maxwell equation becomes
\begin{align}
    \nabla^{b}\nabla_{b}A_{a}&=0,\label{eq:wave_eq_of_A}
    \\\nabla^{a}A_{a}&=0,\label{eq:gc_Lorenz_of_A}
\end{align}
in the leading-order and
\begin{align}
    \nabla^{b}\nabla_{b}X_{a}&=2\nabla_{b}A_{c}\tensor{\dot{C}}{^{c}_{a}^{b}}+h^{bc}\nabla_{b}\nabla_{c}A_{a},\label{eq:wave_eq_of_X}
    \\\nabla^{a}X_{a}&=h^{ab}\nabla_{a}A_{b},\label{eq:gc_Lorenz_of_X}
\end{align}
in the next-to-leading-order where
\begin{equation}
    \tensor{\dot{C}}{^{a}_{bc}}=\frac{1}{2}\left(\nabla_{c}\tensor{h}{^{a}_{b}}+\nabla_{b}\tensor{h}{^{a}_{c}}-\nabla^{a}h_{bc}\right).
\end{equation}

We consider a plane wave solution for \Cref{eq:wave_eq_of_A} given by
\begin{equation}
    A_{a}=\intCl\tilde{A}_{a}\left(l\right)\iq,\label{eq:plane_wave}
\end{equation}
where $\mathcal{C}$ is an equivalence class on $\mathcal{N}$ with respect to the relation $\left\{\left(l,l'\right)\in\mathcal{N}\times\mathcal{N}:l\cdot l'=0\right\}$, $\tilde{A}\,d\mathcal{C}$ is the infinitesimal amplitude, and $Q$ is the phase such that $l^{a}=\nabla^{a}Q$. Note that $\tilde{A}\left(-l\right)=\tilde{A}^{*}\left(l\right)$ because $A$ is real. From the gauge condition in \Cref{eq:gc_Lorenz_of_A}, we get
\begin{equation}
    l\cdot\tilde{A}\left(l\right)=0,
\end{equation}
for all $l\in\mathcal{C}$.

\section{Perturbation of EMWs}

We see that \Cref{eq:wave_eq_of_X} is a wave equation with source terms in its right-hand side. For simple algebra, let us introduce an additional gauge condition for $h$ given by
\begin{equation}
    l^{b}h_{ab}=0,\label{eq:gc_double_null}
\end{equation}
for $l\in\mathcal{C}$. We can prove that this gauge under \Cref{eq:gc_Lorenz_of_h,eq:gc_traceless_of_h} is always possible in a similar way to \citet[sec. 2.1]{flanaganBasicsGravitationalWave2005}. Then, \Cref{eq:wave_eq_of_X} is simplified as
\begin{equation}
    \nabla^{b}\nabla_{b}X_{a}=-\int_{\bar{\mathcal{N}}}d^{3}\bar{\mathcal{N}}\left(k\right)\intCl\left(k\cdot l\right)\tilde{h}_{ab}\tilde{A}^{b}\ip\iq,\label{eq:wave_eq_of_X_with_gc}
\end{equation}
where we restrict the domain of $k$ as $\bar{\mathcal{N}}\equiv\mathcal{N}-\mathcal{C}$ because the integrand vanishes when $k\in\mathcal{C}$.

The general solution for the equation is given by
\begin{equation}
    X_{a}=\Xp_{a}+\Xh_{a},\label{eq:gs_of_X}
\end{equation}
where $X^{\mathrm{p}}$ is a particular solution to match the source terms, and $X^{\mathrm{h}}$ is a homogeneous solution that satisfies the homogeneous wave equation. We consider a form of particular solution given by
\begin{equation}
    \Xp_{a}=\intV\tXp_{a}\left(k,l\right)\ip\iq,
\end{equation}
where $\mathcal{V}=\bar{\mathcal{N}}\times\mathcal{C}$ and $d^{4}\mathcal{V}\left(k,l\right)=d^{3}\bar{\mathcal{N}}\left(k\right)d\mathcal{C}\left(l\right)$. Then, we obtain
\begin{equation}
    \tXp_{a}=\frac{1}{2}\tilde{h}_{ab}\tilde{A}^{b},
\end{equation}
from \Cref{eq:wave_eq_of_X_with_gc}. Because it implies that
\begin{align}
    \nabla^{b}\nabla_{b}\Xh_{a}&=0,
    \\\nabla^{a}\Xh_{a}&=0,
\end{align}
from \Cref{eq:gc_Lorenz_of_X,eq:wave_eq_of_X_with_gc}, we have the wave solution for $\Xh$ as
\begin{equation}
    \Xh_{a}=\intNm\,\tXh_{a}\left(m\right)\ir,\label{eq:Xh}
\end{equation}
where $\tXh\,d^{3}\mathcal{N}$ is the infinitesimal amplitude, and $R$ is the phase such that $m^{a}=\nabla^{a}R$. To determine the amplitude, we need a proper boundary condition. However, it is difficult to impose a physical condition on $X$ because it is not gauge-invariant. Therefore, we first propose gauge-invariant quantities for EMWs and then complete the general solution.

\section{Invariants for EMWs}

We introduce two invariants for an electromagnetic field  defined by
\begin{align}
    I_{1}&=\frac{1}{4}F^{ab}F_{ab},
    \\I_{2}&=\frac{1}{4}F^{ab}\left(\star F\right)_{ab},
\end{align}
where $F_{ab}=2\nabla_{[a}A_{b]}$ is the electromagnetic field, $\star$ is the Hodge star operator such that $\left(\star F\right)_{ab}=\frac{1}{2}\epsilon_{abcd}F^{cd}$, and $\epsilon$ is the Levi-Civita tensor. Note that they do not depend on any observer, and all scalar fields for an electromagnetic field are functions of $\left(I_{1},I_{2}\right)$ if they are constructed only with $g$ and $\epsilon$ \citep{escobarInvariantsElectromagneticField2014}. The invariants can be measured by a simple experiment using a straight conducting wire with a length $\vec{L}$ and a three-velocity $\vec{V}$ in a laboratory. The electromagnetic field $\left(\vec{E},\vec{B}\right)$ in the laboratory induces the electromotive force to the wire as
\begin{equation}
    \mathcal{E}_{\mathrm{emf}}=\left(\vec{E}+\vec{V}\times\vec{B}\right)\cdot\vec{L}.
\end{equation}
From measurements of $\mathcal{E}_{\mathrm{emf}}$ by six wires, each with a independent pair of $\left(\vec{L},\vec{V}\right)$, we can determine $\left(\vec{E},\vec{B}\right)$. Then, the invariants are obtained by
\begin{align}
    I_{1}&=\frac{1}{2}\left(\vec{E}^{2}-\vec{B}^{2}\right),
    \\I_{2}&=\vec{E}\cdot\vec{B}.
\end{align}

Let us return to our case. The perturbed invariants are expanded as
\begin{equation}
    \eps{I}{_{i}}=\epsilon^{2}\left\{I_{i}+\epsilon J_{i}+O\left(\epsilon^{2}\right)\right\},
\end{equation}
where $I_{i}$ are the leading-order quantities, and $J_{i}$ are the perturbations for $i\in\left\{1,2\right\}$. For the plane EMWs given in \Cref{eq:plane_wave}, it can be easily seen that the invariants $I_{i}$ vanish. Therefore, their perturbations $J_{i}$ are gauge-invariant. This fact provides a strong motivation to use the invariants as measurement quantities when observing GWs by the plane EMWs. The perturbations $J_{i}$ are explicitly obtained by
\begin{align}
    J_{1}&=\frac{1}{2}F^{ab}Y_{ab},
    \\J_{2}&=\frac{1}{2}F^{ab}\left(\star Y\right)_{ab},
\end{align}
where $Y_{ab}\equiv2\nabla_{[a}X_{b]}$. Substituting \Cref{eq:gs_of_X} into the above ones, we assort terms as
\begin{equation}
    J_{i}=\Jp_{i}+\Jh_{i},\label{eq:general_solution_of_J}
\end{equation}
where $\Jp_{i}$ and $\Jh_{i}$ are parts with $\Xp$ and $\Xh$, respectively. Then, $\Jp_{i}$ and $\Jh_{i}$ become
\begin{align}
    \Jp_{i}&=F^{ab}\intV\,\left(\tJp_{i}\right)_{ab}\left(k,l\right)\ip\iq,\label{eq:Jp}
    \\\Jh_{i}&=F^{ab}\intNm\left(\tJh_{i}\right)_{ab}\left(m\right)\ir,\label{eq:Jh}
\end{align}
where $\left(\tJp_{1}\right)_{ab}=\frac{1}{2}\ii k_{a}\tilde{h}_{bc}\tilde{A}^{c}$, $\tJp_{2}=\star\tJp_{1}$, and $\tJh_{i}$ are the amplitudes for $\Jh_{i}$.

\section{Plane Boundary Condition}

Let us consider a two-dimensional timelike congruence $\mathcal{P}_{\epsilon}$ in the perturbed spacetime $\mathcal{M}_{\epsilon}$ that is a collection of world lines for EMW emitters. We assume that the emitters in $\mathcal{P}_{\epsilon}$ keep $I_{i}$ vanishing as $\left(I_{i}\right)_{\mathcal{P}_{\epsilon}}=0$. For a one-parameter group of diffeomorphism $\phi_{\epsilon}:\mathcal{M}_{0}\rightarrow\mathcal{M}_{\epsilon}$ such that $\phi_{\epsilon}\left[\mathcal{P}_{0}\right]=\mathcal{P}_{\epsilon}$, we obtain $\left(\mathcal{L}_{\upsilon}I_{i}\right)_{\mathcal{P}_{0}}=0$ where $\upsilon$ is the generator of $\phi_{\epsilon}$. These conditions hold for any gauge due to the gauge-invariances of $J_{i}$. Therefore, we conclude that
\begin{equation}
    \left(J_{i}\right)_{\mathcal{P}_{0}}=0.\label{eq:bc_of_J}
\end{equation}

\begin{figure}
    \centering
    \includegraphics[width=0.7\columnwidth]{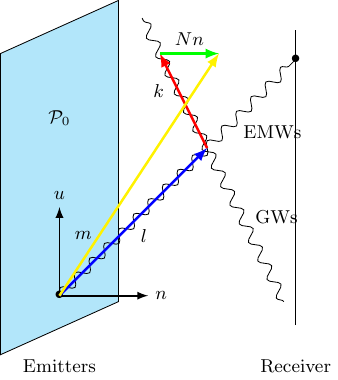}
    \caption{Emitters and a receiver of EMWs are described by the timelike plane $\mathcal{P}_{0}$ orthogonal to $n$ and the world line parallel to $u$ in the spacetime diagram. Vectors in the diagram illustrate determination of $m$, which is the wave vector for $\Jh_{i}$, to meet the boundary condition at $\mathcal{P}_{0}$ when wave vectors $k$ and $l$ for GWs and EMWs, respectively, are given. The value of $N$ is chosen such that $m$ is null and have a causal direction from the emitters to the receiver.}
    \label{fig:emitters_and_receiver}
\end{figure}

We assume that $\mathcal{P}_{0}$ in $\mathcal{M}_{0}$ forms a three-dimensional timelike plane with a unit normal vector field $n$ toward a receiver, and the emitters on $\mathcal{P}_{0}$ are in geodesic motion with a four-velocity vector field $u$ such that $u$ and $n$ are constant over the plane and orthogonal to each other as in \Cref{fig:emitters_and_receiver}. Also, the emitters emit the plane EMWs as in \Cref{eq:plane_wave} with $\mathcal{C}=\left\{l:l\cdot\left(u+n\right)=0\right\}$. Substituting \Cref{eq:Jp,eq:Jh} into \Cref{eq:bc_of_J}, we get the boundary condition given by
\begin{multline}
    \left[F^{ab}\left\{\intV\left(\tJp_{i}\right)_{ab}\left(k,l\right)\ip\iq\right.\right.
    \\\left.\left.+\int_{\mathcal{N}_{\mathrm{c}}}d^{3}\mathcal{N}_{\mathrm{c}}\left(m\right)\,\left(\tJh_{i}\right)_{ab}\left(m\right)\ir\right\}\right]_{\mathcal{P}_{0}}=0,
\end{multline}
where we restrict the domain of $m$ to causal directions to the receiver as $\mathcal{N}_{\mathrm{c}}=\left\{m\in\mathcal{N}^{+}:m\cdot n>0\right\}\cup\left\{m\in\mathcal{N}^{-}:m\cdot n<0\right\}$. To obtain $\Jh_{i}$ from the above, we have to find $R$ that matches $P+Q$ on $\mathcal{P}_{0}$. 

The phase matching condition is given by
\begin{equation}
    D\left[\mathcal{P}_{0};k,l\right]=0,
\end{equation}
where $D\left(;k,l\right)\equiv R\left(;k,l\right)-P\left(;k\right)-Q\left(;l\right)$ is the phase difference of $\left(k,l\right)$ oscillation term in $\Jh_{i}$ from that of $\Jp_{i}$. The matching condition implies
\begin{equation}
    m^{a}-k^{a}-l^{a}=Nn^{a},
\end{equation}
where $m^{a}-k^{a}-l^{a}=\nabla^{a}D$, and $N$ is a constant. The null condition of $m$ determines
\begin{equation}
    N\left(k,l\right)=-\left(k+l\right)\cdot n\pm\sqrt{\left\{\left(k+l\right)\cdot n\right\}^{2}-2\left(k\cdot l\right)},
\end{equation}
where we have to choose plus sign when $m$ is future-directed and minus sign when $m$ is past-directed because we restricted $m\in\mathcal{N}_{\mathrm{c}}$. An illustration of determining $m$ is given in \Cref{fig:emitters_and_receiver}. With the solution $D$ of the equation $\nabla D=N n$, we construct the homogeneous solution as
\begin{equation}
    \Jh_{i}=-F^{ab}\intV\left(\tJp_{i}\right)_{ab}\left(k,l\right)\id\ip\iq.\label{eq:homogeneous_solution_of_J}
\end{equation}
Collecting \Cref{eq:Jp,eq:homogeneous_solution_of_J} into \Cref{eq:general_solution_of_J}, we get the general solution of $J_{i}$ as
\begin{equation}
    J_{i}=F^{ab}\intV\left(\tJp_{i}\right)_{ab}\left(k,l\right)\left(1-\id\right)\ip\iq.\label{eq:general_solution_of_J_value}
\end{equation}
Note that the general solution satisfies the boundary condition in \Cref{eq:bc_of_J}.

To interpret our results, we decompose $k$ and $l$ into
\begin{align}
    k&=\fg\left(u+\kappa\right),
    \\l&=\fe\left(u+n\right),
\end{align}
where $\fg\equiv-k\cdot u$ is the frequency of GWs, $\kappa\equiv k/\fg-u$ is the spatial unit vector of GW propagation, $\fe\equiv-l\cdot u$ is the frequency of EMWs. Then, $N$ becomes
\begin{multline}
    N\left(k,l\right)=-\fg\cos\theta-\fe
    \\\pm\sqrt{\left\{\fe+\fg\left(1+\sin\theta\right)\right\}\left\{\fe+\fg\left(1-\sin\theta\right)\right\}},
    \label{eq:N_in_coord}
\end{multline}
where $\cos\theta=\kappa\cdot n$. By introducing a global inertial coordinate system $\left\{t,\vec{x}\right\}$ such that $\left(\partial\middle/\partial t\right)^{a}=u^{a}$ and assuming the world line of $\vec{x}=0$ belongs to $\mathcal{P}_{0}$, we get
\begin{equation}
    D\left(\vec{x};k,l\right)=N\left(k,l\right)\left(n\cdot\vec{x}\right).\label{eq:D_in_coord}
\end{equation}

\begin{figure}
    \centering
    \includegraphics[width=0.9\columnwidth]{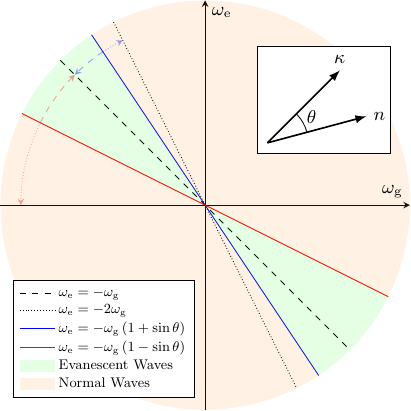}
    \caption{We show regions of normal (orange) and evanescent (green) waves for oscillation terms in $\Jh_{i}$ on the plane of $\fg$ and $\fe$. The boundaries between two regions are determined by $\fe=-\fg\left(1+\sin\theta\right)$ (blue) and $\fe=-\fg\left(1-\sin\theta\right)$ (red) line, where $\theta$ is the angle between the spatial directions of GWs and EMWs with respect to the emitters. When $\theta=\pi/2$, the lines become $\fe=-2\fg$ (dotted) and $\fe=0$, respectively. When $\theta\rightarrow0$, the lines converge to $\fe=-\fg$ (dashed) line.}
    \label{fig:frequency_region}
\end{figure}

An oscillation term of $\left(k,l\right)$ in $\Jh_{i}$ such that the inside of root in \Cref{eq:N_in_coord} is negative means evanescent waves although the complex extension of $\mathcal{N}$ in \Cref{eq:Xh} is required to embrace them. In that case, we need plus sign in \Cref{eq:N_in_coord} to get a physical solution that decays toward the direction of $n$. Note that the evanescent oscillation term depends on the spatial position of emitters through \Cref{eq:D_in_coord}, but it is diluted where $n\cdot\vec{x}$ is large enough. We explore regions of normal and evanescent waves on the parameter space of $\left(\fg,\fe\right)$ in \Cref{fig:frequency_region}.

When $\left|\fe\right|\gg\left|\fg\right|$, which is the geometrical optics limit, $N\rightarrow\fg\left(1-\kappa\cdot n\right)$ and $P+D$ in \Cref{eq:general_solution_of_J_value} becomes
\begin{equation}
    \left(P+D\right)\left(t,\vec{x};k,l\right)=P\left(t-n\cdot\vec{x},\vec{x}-n\left(n\cdot\vec{x}\right);k\right),
\end{equation}
which is the retarded phase of GWs from $\mathcal{P}_{0}$. It is consistent with the analyses on the pulsar timing based on the geometrical optics in \citet[sec. 3.2]{yunesGravitationalWaveTestsGeneral2013} and \citet[chap. 23.1]{Maggiore2018}. For GWs and EMWs having a geometrical-optical separation of frequency domains, we have simple forms for $J_{i}$ as
\begin{align}
    J_{1}&=\frac{1}{2}F^{ab}A^{c}\left(1-\mathfrak{D}\right)\nabla_{a}h_{bc},
    \\J_{2}&=\frac{1}{2}\left(\star F\right)^{ab}A^{c}\left(1-\mathfrak{D}\right)\nabla_{a}h_{bc},
\end{align}
where $\mathfrak{D}$ is the retardation operator defined by
\begin{equation}
    \mathfrak{D}\left\{Z\left(t,\vec{x}\right)\right\}=Z\left(t-n\cdot\vec{x},\vec{x}-n\left(n\cdot\vec{x}\right)\right),
\end{equation}
for a tensor field $Z$.

\section{Antenna Patterns for GWs}

To see the nature of EMW perturbation, let us consider a simple case with monochromatic GWs and EMWs as
\begin{align}
    h_{ab}&=2\Re\left\{\tilde{h}_{ab}\ip\right\},
    \\A_{a}&=2\Re\left\{\tilde{A}_{a}\iq\right\},
\end{align}
such that $k^{a}=\nabla^{a}P$ and $l^{a}=\nabla^{a}Q$ are future-directed. Substituting them into \Cref{eq:general_solution_of_J_value}, we identify that both waves make three oscillation terms in
\begin{equation}
    J_{i}=2\Re\left\{\sum_{j=-1}^{1}\tilde{J}_{i}^{\mathrm{p}\left(j\right)}\tilde{D}^{\left(j\right)}e^{\ii\left(P+2jQ\right)}\right\},
\end{equation}
where $\tilde{J}_{i}^{\mathrm{p}\left(j\right)}$ are the amplitudes that come from $\tilde{J}_{i}^{\mathrm{p}}$ in \Cref{eq:general_solution_of_J_value} and
\begin{equation}
    \tilde{D}^{\left(j\right)}\equiv1-\frac{1+j}{2}e^{\ii D\left(;k,+l\right)}-\frac{1-j}{2}e^{\ii D\left(;k,-l\right)}.
\end{equation}
Note that $\tilde{D}^{\left(j\right)}$ are just factors determined by a spatial position of the receiver relative to the emitters.

We introduce an adapted observer for both waves with four-velocity $v$ and its unit spatial vector $\lambda$ defined by
\begin{align}
    v^{a}&=\frac{1}{2\omega}\left(l^{a}+k^{a}\right),
    \\\lambda^{a}&=\frac{1}{2\omega}\left(l^{a}-k^{a}\right),
\end{align}
where $\omega\equiv\sqrt{-\left(k\cdot l\right)/2}$. The observer sees that the EMWs and GWs are propagating toward $+\lambda$ and $-\lambda$, respectively, and have the same frequency $\omega$ as shown by
\begin{align}
    l^{a}&=\omega\left(v^{a}+\lambda^{a}\right),
    \\k^{a}&=\omega\left(v^{a}-\lambda^{a}\right).
\end{align}
In addition, \Cref{eq:gc_Lorenz_of_h,eq:gc_double_null} of our gauge conditions imply
\begin{align}
    v^{b}h_{ab}&=0,
    \\\lambda^{b}h_{ab}&=0.
\end{align}
The gauge we chose was actually the transverse-traceless gauge in the adapted observer.

Imposing the radiation gauge as
\begin{equation}
    v\cdot\tilde{A}=0,
\end{equation}
we have a form of $\tilde{A}$ given by
\begin{equation}
    \tilde{A}_{a}=\frac{1}{\ii\omega}E\left(\cos\chi\,p_{a}+\ii\sin\chi\,s_{a}\right)e^{\ii\delta_{\mathrm{e}}},
\end{equation}
where $E>0$ is the real amplitude of the electric field in the adapted observer, $-\pi/4\leq\chi\leq\pi/4$ is the ellipticity of the polarization ellipse, $p$ and $s$ are the unit spatial vectors parallel to the major and minor axes of the polarization ellipse, respectively, such that $\left\{v,p,s,\lambda\right\}$ forms a right-handed orthonormal basis, and $\delta_{\mathrm{e}}$ is a constant phase to realize this form. The method to orthogonalize real and imaginary parts by adjusting a constant phase was introduced in \citet[chap. 6]{landauClassicalTheoryFields1975}. By the analogy of the method, we get the form of $\tilde{h}$ as
\begin{equation}
    \tilde{h}_{ab}=H\left(\cos\eta\,e^{+}_{ab}+\ii\sin\eta\,e^{\times}_{ab}\right)e^{\ii\delta_{\mathrm{g}}},
\end{equation}
where $H>0$ is the real amplitude of the GWs, $-\pi/4\leq\eta\leq\pi/4$ is the ellipticity, and $\delta_{\mathrm{g}}$ is a constant phase to realize this form. The orthonormal basis $\left\{e^{+},e^{\times}\right\}$ is given by
\begin{align}
    e^{+}_{ab}&=\frac{1}{\sqrt{2}}\left(x_{a}x_{b}-y_{a}y_{b}\right),
    \\e^{\times}_{ab}&=\frac{1}{\sqrt{2}}\left(x_{a}y_{b}+y_{a}x_{b}\right),
\end{align}
where $x$ and $y$ are the unit spatial vectors such that $\left\{v,x,y,-\lambda\right\}$ forms a right-handed orthonormal basis. Note that $\left\{x,y\right\}$ and $\left\{p,s\right\}$ span an identical vector space orthogonal to $\left\{v,\lambda\right\}$. Therefore, we define a polarization angle $\xi$ of the GWs as an angle of $x$ with respect to frame $\left\{p,s\right\}$ such that $\cos\xi=p\cdot x$ and $\sin\xi=s\cdot x$.

\begin{figure}
    \centering
   \includegraphics[width=0.9\columnwidth]{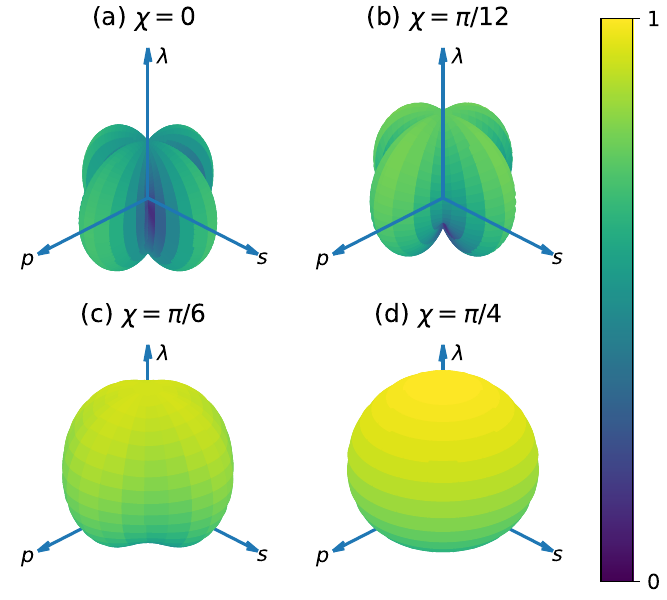}
    \caption{We plot the pattern functions $\mathfrak{F}_{1}^{\left(+1\right)}\left(\chi,\eta,\xi\right)$ for $\chi\in\left\{0,\pi/12,\pi/6,\pi/4\right\}$ on the frame $\left\{p,s,\lambda\right\}$ with azimuthal angle $\vartheta=\pi/2-2\eta$ and polar angle $\varphi=\xi$. Note that the GWs are propagating to $-\lambda$ in the adapted observer, not the direction of $\left(\vartheta,\varphi\right)$. We utilize the azimuthal angle $\vartheta$ and polar angle $\varphi$ to present the ellipticity and the polarization angle of GWs, respectively. When $\vartheta$ is $0$, $\pi/2$, and $\pi$, the GWs exhibit in the right circular, linear, and left circular polarization, respectively.}
    \label{fig:antenna}
\end{figure}

As a result, $J_{i}$ become
\begin{equation}
    J_{i}=2E^{2}H\sum_{j=-1}^{1}\left|\tilde{D}^{\left(j\right)}\right|\mathfrak{F}_{i}^{\left(j\right)}\cos\left(P+2jQ+\delta^{\left(j\right)}_{i}\right),
\end{equation}
where $\delta^{\left(j\right)}_{i}$ are phase constants and $\mathfrak{F}_{i}^{\left(j\right)}\left(\chi,\eta,\xi\right)$ are pattern functions given by
\begin{align}
    \mathfrak{F}_{i}^{\left(0\right)}&=\cos\left(2\chi\right)\sqrt{1+\cos\left(2\eta\right) \cos\left(4\xi+\left(i-1\right)\pi\right)},
    \\\mathfrak{F}_{i}^{\left(\pm1\right)}&=\frac{1}{2}\sqrt{\left(\mathfrak{F}_{i}^{\left(0\right)}\right)^{2}+2\sin\left(2\chi\right)\left\{\sin\left(2\chi\right)\pm\sin\left(2\eta\right)\right\}}.
\end{align}
We show plots of $\mathfrak{F}_{1}^{\left(+1\right)}$ for different $\chi>0$ in \Cref{fig:antenna} that cover all representative cases because other pattern functions are given by
\begin{align}
    \mathfrak{F}_{2}^{\left(j\right)}\left(\chi,\eta,\xi\right)&=\mathfrak{F}_{1}^{\left(j\right)}\left(\chi,\eta,\xi+\pi/4\right),
    \\\mathfrak{F}_{i}^{\left(-1\right)}\left(\chi,\eta,\xi\right)&=\mathfrak{F}_{i}^{\left(+1\right)}\left(\chi,-\eta,\xi\right),
    \\\mathfrak{F}_{i}^{\left(0\right)}\left(\chi,\eta,\xi\right)&=2\cos\left(2\chi\right)\mathfrak{F}_{i}^{\left(\pm1\right)}\left(0,\eta,\xi\right),
    \\\mathfrak{F}_{i}^{\left(j\right)}\left(-\chi,\eta,\xi\right)&=\mathfrak{F}_{i}^{\left(j\right)}\left(\chi,-\eta,\xi\right).
\end{align}
Defining the overall gain as
\begin{equation}
    \mathfrak{F}\equiv\sqrt{\sum_{i=1}^{2}\sum_{j=-1}^{1}\left(\mathfrak{F}_{i}^{\left(j\right)}\right)^{2}}=\sqrt{2+\cos^{2}\left(2\chi\right)},
\end{equation}
we find that the measurement with EMWs in the linear polarization ($\chi=0$) has the most strong gain of GWs. However, in this case, we cannot simultaneously determine the ellipticity $\eta$ and the polarization angle $\xi$ of GWs. For the measurement with EMWs in the circular polarization ($\chi=\pi/4$), the polarization angle $\xi$ of GWs is not determined.

\section{Summary and Discussions}

We have solved the perturbed Maxwell equation for the perturbation of electromagnetic potential assuming plane EMWs at the background and general GWs. By introducing an orthogonality condition of the GW amplitude to the wave vector of EMW, we can considerably simplify the equation. Because the equation is a wave equation with source terms, the general solution is the superposition of the particular solution that is given by the source terms and the homogeneous solution that is determined by boundary conditions. Note that the perturbation of electromagnetic potential is not gauge-invariant. To give physical boundary conditions, a gauge-invariant quantity is required.

Electromagnetic invariants have many advantages in terms of measurement. First, they are independent of the frames established by detectors. It means that we do not need to consider the frame perturbation induced by GWs. Second, the perturbations of electromagnetic invariants are gauge-invariant because their background value vanishes. Therefore, they are observables that can be measured without any ambiguity. Moreover, we can impose physical boundary conditions using the invariants. Third, the fact that the background values of the invariants are zero is also one of the advantages. In principle, it is easier to measure a perturbation of zero value than one of nonzero value.

The perturbations of invariants are superposed by the particular and homogeneous parts that come from the particular and homogeneous solution, respectively. To get the homogeneous part, we imposed the boundary condition that the perturbations of invariants vanish on the three-dimensional timelike plane that is a collection of geodesic word lines for EMW emitters. It determines the homogeneous part that has a phase difference factor from the particular part. Some oscillation components in the homogeneous part are evanescent waves decaying to the receiver. These components might be useful because information on the position of emitters is attenuated at a great distance. In the geometrical optics limit, the GWs in the homogeneous part become the retarded GWs from the emitters, which is consistent with the literature on pulsar timing.

To see the nature of EMW perturbation, we have considered the simple case of monochromatic GWs and EMWs and introduced the adapted observer for both waves. In the observer, GWs and EMWs have an identical frequency and propagate to opposite directions. Because amplitudes of GWs and EMWs in our gauge live in the identical tensor space orthogonal to the propagation directions and four-velocity of the adapted observer, we can define the polarization angle as the tilt angle between polarization ellipses of both waves. Then, the pattern functions of GWs in the perturbations of invariants are provided as the functions of the polarization angle and ellipticities of GWs and EMWs. The measurement with EMWs in the linear polarization has the most strong gain of GWs although the ellipticity and the polarization angle of GWs are not determined simultaneously.

\begin{acknowledgments}
C.P. thanks Yeong-Bok Bae for his valuable comments and APCTP for its hospitality during the completion of this work. This work was supported by IBS under the project code, IBS-R018-D1.
\end{acknowledgments}

\bibliographystyle{aasjournal}
\bibliography{references}

\end{document}